\documentclass[preprint,aps,amsmath,showpacs]{revtex4}
\usepackage{graphics}

\begin{document}

\title{Robust control of decoherence\\ in realistic one-qubit quantum gates}
\author{V. Protopopescu}
\email{protopopesva@ornl.gov}
\author{R. Perez}
\author{C. D'Helon}
\author{J. Schmulen}
\affiliation{Center for Engineering Science Advanced Research, Computer Science and Mathematics Division,\\ Oak Ridge National Laboratory, Oak Ridge, TN 37831-6355 USA}

\begin{abstract}
We present an open loop (bang-bang) scheme to control decoherence in a generic one-qubit quantum gate and implement it in a realistic simulation. The system is consistently described within the spin-boson model, with interactions accounting for both adiabatic and thermal decoherence. The external control is included from the beginning in the Hamiltonian as an independent interaction term. After tracing out the environment modes, reduced equations are obtained for the two-level system in which the effects of both decoherence and external control appear explicitly. The controls are determined exactly from the condition to eliminate decoherence, i.e. to restore unitarity. Numerical simulations show excellent performance and robustness of the proposed control scheme.
\end{abstract}

\maketitle

\newpage
\section{Introduction}
\label{Introduction}

Quantum computing holds an extremely appealing practical promise and exerts a no less remarkable intellectual fascination.  Indeed, the quantum algorithms proposed until now could simplify considerably the computational complexity of classically hard problems.  While their actual implementation in robust and realistically sized quantum computers remains still a promise, there is general consensus in expecting further significant advances both in developing new quantum algorithms and the hardware they could be implemented in \cite {Pitt, NC}.  

In the presently accepted paradigm, a quantum computation is a sequence of unitary transformations effected by the quantum computer upon a state in a finite dimensional Hilbert (i.e. a complex Euclidian) space.  At the end of the computation, the system is left in a state which, upon measurement (read-out), would yield with high probability the desired result.  While the read-out is non-invertible, quantum computation \textit{per se} is deterministic and reversible.  Thus, the realization of any quantum computer relies on maintaining \textit{quantum coherence} in the system, for a period of time spanning at least the duration of the desired computation.  Unfortunately, due to coupling to the environment which is not explicitly accounted for in quantum computation, the evolution within the quantum computer loses its unitary character; in other words, the system decoheres \cite{Pitt, NC, Gardiner91, Giulini, BEZ, Weiss99}.

Any quantum computation, no matter how general or convoluted, can be approximated to arbitrary precision using a universal set of quantum gates e.g., the discrete set of gates made up of the CNOT, Hadamard, phase, and $\pi/8$ gates \cite{Pitt, NC}. Thus, a successful approach to the decoherence problem in any quantum computer requires to address and solve the problem at the elementary gate level.  To be specific, in this paper we shall focus on the NOT gate, which will serve as an illustration for any one-qubit gate. The analysis of the two-qubit CNOT gate will be considered elsewhere.

The NOT gate acts on a single qubit, which is the most general superposition state of a two-level system. Since the set containing the identity and the Pauli matrices, $\{I, \sigma_x, \sigma_y, \sigma_z \}$, forms a basis in the space of $2 \times 2$ matrices, the coupling of a two-level system to the outside world can be essentially realized through the matrices $\sigma_z$ and $\sigma_x$. These two couplings lead to different effects, known as \textit{adiabatic} and \textit{thermal} decoherence, respectively.

In general, quantum decoherence is a very fast process, whose speed depends primarily on the size and temperature of the computer, but may depend significantly on other factors as well, such as imperfect state preparation, undesired excitations, inaccuracies in manipulations of the logic gates, explicit stochasticity and noise, etc.  For example, decoherence is induced by the spontaneous emission of a photon from a two-level atom, where the environment is represented by the weakly-coupled electromagnetic field. Whereas the evolution of the two-level system and its environment considered together as a large (closed) quantum system is in principle unitary, the subdynamics of the two-level system generally is not. Clearly, this behavior represents a major stumbling block on the path to realizing a usable quantum computer at room temperature.

Several approaches have been proposed to eliminate or mitigate the undesirable effects of decoherence in open quantum systems, including open loop (quantum bang-bang) control \cite{Tombesi_ol,Viola98,Viola99a,Viola99b}, decoherence free subspaces (DFS) \cite{Zanardi97,Lidar98}, quantum error correction \cite{Preskill98,Knill00} and quantum feedback \cite{Tombesi_qf,Wang01}.

The open loop (quantum bang-bang) control has been pioneered by the seminal work of Lloyd and Viola (see Refs. \cite{Viola98,Viola99a,Viola99b} and references therein). Vitali and Tombesi \cite{Tombesi_ol} have also considered applying a sequence of frequent parity kicks, as well as an appropriate stochastic modulation, in order to control decoherence. Essentially this type of approach relies on applying rapid control pulses to the system in order to dynamically cancel the system-environment interaction. It has been shown that decoherence can be effectively suppressed if the pulse rate is much higher than the frequency corresponding to the correlation time of the environment. Thus, this open-loop control technique "freezes" decoherence in a manner analogous to the quantum Zeno effect.

DFS are special subspaces of the Hilbert space, which are inherently immune to decoherence, due to specific symmetries in the system-environment interaction. This concept was first introduced by Zanardi and Rasetti \cite{Zanardi97}, and has been recently generalized to noiseless subsystems \cite{Knill00}. It has also been shown \cite{Viola00,WL} that bang-bang control can be used to create the conditions required for large DFS. The successful use of decoherence free subspaces (DFS) to passively stabilize a quantum state has been reported for photon singlets \cite{Kwiat00}, for a pair of trapped ions \cite{Kielpinski01}, and more recently for a noiseless subsystem via liquid state NMR \cite{Viola01}.

Quantum error correction is historically the first method proposed to counter decoherence, and also the only one to date for which an accuracy threshold theorem has been proved \cite{Knill96}. It is essentially an active form of quantum feedback control implemented on a system encoding redundant information. Valid states of the system are restricted to special Hilbert subspaces (codes) so that any loss of information due to decoherence (errors) can be diagnosed and reversed. The system is closely monitored to observe errors, which can be corrected later by conditionally performing appropriate feedback operations. It has been shown that fault-tolerant recovery which allows for errors in the correcting feedback, if the error rate is below a critical value.

The efficient implementation of both DFS and quantum error correction techniques requires a large amount of ancillary quantum memory. Thus quantum feedback (closed-loop) and open-loop control strategies have the advantage that in general they do not require any ancillary quantum resources.

Quantum feedback can also be used to stabilize a quantum state against decoherence, as first shown by Tombesi \textit{et al.} in a series of papers \cite{Tombesi_qf}. For example, in a scheme recently proposed by Wang {\it et al.} \cite{Wang01} for a two-level atom, photocurrent feedback from homodyne detection of spontaneous emission is used to alter the atom-environment dynamics. The quantum state of the atom is driven to and then remains in a pure target state, even though the atom is still spontaneously emitting. Of course, the feedback delay time has to be shorter than the time scale corresponding to the spontaneous emission rate, in order to successfully maintain the atom in the target state. We note that this quantum feedback scheme cannot be used to maintain an unknown state such as that obtained from a quantum computation, it can only drive the system to a known target state.

In this paper, an open-loop control scheme is proposed and simulated, to eliminate the effects of decoherence in a two-level system. While the general philosophy is following in the steps of Lloyd and Viola's seminal results \cite{Viola98,Viola99a,Viola99b}, our implementation has several new aspects. More precisely, we present the first implementation of a quantum control scheme in which: (i) decoherence and control are taken to act simultaneously within a realistic model, which allows one to deal with thermal decoherence; (ii) the required control is directly related to and calculated from the decoherence effects, which presents the practical advantage of maintaining the frequency and amplitude of the required controls at manageable levels; and (iii) the effect of imperfect control pulses is assessed. We are aware that determining the decoherence behavior of a quantum system can be difficult in practice, but if we know it then our control scheme can take advantage of this, to tailor more efficient controls. 

The remaining of the paper is organized as follows: In section 2 we present the spin-boson model for a two-level system in contact with the environment and an external control.  In Section 3 we calculate the reduced density matrices for the two level system after tracing out the environment modes, in the case of adiabatic and thermal decoherence.  The control strategy is implemented in Section 4 and the results are discussed in Section 5.


\section{Spin-Boson Model}
\label{Spin-Boson Model}
The exact quantum mechanical description of the system starts from the complete Hamiltonian that accounts for the two-dimensional system and the environment which, in principle, is the rest of the Universe.  In different situations the environment  predominantly manifests itself in different ways, e.g., electromagnetic modes, acoustic modes, etc.  Thus, in many instances, a reasonable and generic model is represented as a system of noninteracting boson modes.  The two-level system interacting with these modes is known as the spin-boson model \cite{Weiss99}.  We write the complete Hamiltonian in the form:

\begin{equation}
H = H_s + H_e + H_i + H_c
\end{equation}
where 
\begin{equation}
H_s = \sum_{i=1}^2P_{ii}E_i
\end{equation}
and 
\begin{equation}
H_e = \hbar\sum_{q=1}^{\infty}\omega_{0q}a^\dagger_q a_q 
\end{equation}
represent the two-level system and the boson bath. Here $P_{ii}$ denote the projection operators $P_{ii}=|i\rangle \langle i|$, and $E_i$ are the corresponding energies.  The terms  
\begin{equation}
H_i = -\hbar\epsilon(\alpha\sigma_x + \beta\sigma_z)\sum_{q=1}^{\infty}(\Omega_q^\star a_q + \Omega_q a^\dagger_q)
\end{equation}
and 
\begin{equation}
H_c = -\hbar\Omega_F V(t) \sigma_x \cos(\omega_0 t)
\end{equation}
describe the interaction of the two level system with the environment and the external control, $V(t)$, respectively.  The interaction with the environment is parametrized by $\epsilon$ (whose magnitude indicates the strength of the coupling) and can result in a phase flip or a bit flip (a bit-phase flip can also be obtained either by combining the two effects or by including a $\sigma_y$ term).  However, from a formal viewpoint, the effects of $\sigma_x$ and $\sigma_y$ are essentially similar and we shall not consider the latter.  On the other hand, the control which is applied as a fast laser pulse, is assumed to act strongly only upon the spin, without significantly affecting the environment, either directly or indirectly.  

While the spin-boson model in various guises has been thoroughly studied  \cite{Leggett87,Kampen95,Palma96,Mozyrsky98}, the consistent inclusion of the control term, $H_c$, in the Hamiltonian, and the general derivation of the corresponding reduced description are new.

The dynamics of the two-level system considered in this paper is
expressed in terms of parameters normalized to the Rabi frequency, $\Omega_F$, 
namely the time evolution is described in terms of dimensionless Rabi time units,   $\tau=\Omega_F t$, and all the frequencies, namely $\omega_q=\omega_{oq}/\Omega_F$, $\omega=\omega_o/\Omega_F$, $g_q=\Omega_q/\Omega_F$ and $\omega_{12}=\Omega_{12}/\Omega_F$ are also renormalized with respect to the Rabi frequency. The frequency $\Omega_{12}$ is the transition frequency between the energy levels of the two-level system.

The state of the system at $\tau = 0$ is described by the density matrix:

\begin{equation}
\rho(s,e,0)=\rho(s,0)\otimes\rho(e,0)
\end{equation}
where
\begin{equation}
\rho(e,0)=\prod_q{\rho_q(e,0)}
\end{equation}
\begin{equation}
\rho_q(e,0)=[1-e^{-\frac{\hbar\omega_{oq}}{kT}}]^{-1}\sum_{n_q}e^{-\frac{\hbar\omega_{oq}}{kT}n_q}|n_q\rangle\langle n_q|
\end{equation}
and
\begin{equation}
\rho(s,0)=\sum_{i,j=1}^2 \rho_{ij}(0)P_{ij}.
\end{equation}

In general, the decoherence experienced by a two-level system has two distinct timescales, corresponding to the adiabatic and thermal regimes separately.

Adiabatic decoherence is responsible for the decay of transverse $(x,y)$ polarization (the off-diagonal density matrix elements). It acts on a relatively short timescale, such that a superposition state will typically decay to a mixture of states. For larger, macroscopic sub-systems, adiabatic decoherence ensures that quantum superpositions of distinct states will not be observed. The mixture of states generated by adiabatic decoherence will then continue to relax to a stationary value over a longer thermal timescale.

Thermal decoherence generally causes a change of all density matrix elements, and leads to the exponential decrease of the excited state population of the two-level system, due to energy exchanges between the two-level system and its environment. It is commonly neglected in discussing decoherence of quantum systems since its timescale is longer than the timescale for adiabatic decoherence, which is deemed responsible for eliminating most quantum effects \cite{ZHP}. We note that often thermal decoherence is refered to as (quantum) dissipation, and adiabatic decoherence is simply refered to as decoherence.

To simplify the calculations and render them more transparent, it is convenient to write the evolution of the system in the interaction representation, and consider separately the adiabatic ($\alpha=0$, $\beta=1$) and thermal ($\alpha=1$, $\beta=0$) decoherence regimes. A combined treatment of these regimes is straightforward, though computationally cumbersome. The unitary evolution operator in the interaction representation, $U_I$, is related to the unitary evolution operator in the Schr\"odinger representation, $U$, by

\begin{equation}
U_I(s,e,\tau)=e^{i H_0 \tau /\hbar} U(s,e,\tau)
\end{equation}
where $H_0=H_s+H_e$.

Thus the evolution of the density matrix of the system and environment is
\begin{equation}
\frac{\partial\rho}{\partial\tau}=-\frac{i}{\hbar}[H_I,\rho]
\end{equation}
where the interaction Hamiltonian contains both the interaction with the environment and with the applied control:
\begin{equation}
H_I=H_{Ic}+H_{Ii},
\end{equation}
\begin{equation}
H_{Ic}=-\frac{\hbar}{2}V(\tau)\sigma_x,
\end{equation}
\begin{equation}
H_{Ii}=
\begin{cases}
-\epsilon\hbar\sigma_z \sum_{q}(g_q^\star a_q e^{-i\omega_q \tau} + g_q a^\dagger_q e^{i\omega_q \tau})&\text{$\alpha=0$, $\beta=1$} \\
-\epsilon\hbar [ \frac{\sigma_x + i \sigma_y}{2} \sum_q g_q^\star a_q e^{-i(\omega_{12}-\omega_q)\tau} + \frac{\sigma_x - i \sigma_y}{2} \sum_q g_q a^\dagger_q e^{i(\omega_{12}-\omega_q)\tau}) ]&\text{$\alpha=1$, $\beta=0$}
\end{cases}
\end{equation}
These interaction-picture Hamiltonians have been calculated by using the rotating wave approximation, and assuming a zero detuning $\delta=\omega_{12}-\omega=0$.

The evolution equation for $\rho$ admits the formal solution 
\begin{equation}
\rho(s,e,\tau)=U_I(s,e,\tau)\rho(s,e,0)U_I^{\dagger}(s,e,\tau)
\end{equation}
where the evolution operator satisfies the equation
\begin{equation}
\frac{dU_I(s,e,\tau)}{d\tau}=-\frac{i}{\hbar}[H_{Ic}+H_{Ii}]U_I(s,e,\tau)
\end{equation}
given the initial condition $U_I(s,e,0)=1$. The formal solution of this equation is \cite{Scully_qo},
\begin{equation}
U_I(s,e,\tau)={\mathcal T}[\exp\{-\frac{i}{\hbar}\int_0^\tau d\tau^\prime (H_{Ic}(\tau^\prime)+H_{Ii}(\tau^\prime))\}],
\end{equation}
where ${\mathcal T[\hspace{5pt}]}$ represents a time-ordering operator.

Since the various terms in the interaction Hamiltonian $H_I$ do not commute, the complete evolution operator, $U_I(s,e,\tau)$, cannot be calculated exactly, except in special cases. For instance, Viola and Lloyd \cite{Viola98} have shown that if the interaction with the environment $H_{Ii}$ contains only $\sigma_z$, the evolution operator $U_I(s,e,\tau)$ can be substantially simplified by using $\pi$-pulses for the control. However this type of control would not allow the same simplification if the interaction with the environment $H_{Ii}$ contains $\sigma_x$ or $\sigma_y$.

The general Baker-Hausdorff theorem \cite{Gardiner_qn} can be used to expand the evolution operator into an infinite product of exponentials,
\begin{equation}
U_I(s,e,\tau)=e^{-\frac{i}{\hbar}\int_0^\tau dt H_{Ic}(t)} \times e^{-\frac{i}{\hbar}\int_0^\tau dt H_{Ii}(t)} \times e^{-(\frac{i}{\hbar})^2\int_0^\tau dt \int_0^t dt^\prime [H_{Ic}(t),H_{Ii}(t^\prime)]} \times \dots
\end{equation}
We assume that the effect of the control and interaction Hamiltonians is relatively small at all times. Neglecting the commutators, the evolution operator can be written as a first order approximation in the magnitude of the control pulses, $V(\tau)$, and the coupling strength parameter, $\epsilon$, of the system-environment interaction,
\begin{equation}
U_I(s,e,\tau) \approx 
\begin{cases}
exp\{-\frac{i}{\hbar}\int_0^\tau dt H_{Ic}(t)\} \times exp\{-\frac{i}{\hbar}\int_0^\tau dt H_{Ii}(t)\} & \text{$\alpha=0$, $\beta=1$} \\
exp\{-\frac{i}{\hbar}\int_0^\tau dt (H_{Ic}(t)+H_{Ii_x}(t))\} \times exp\{-\frac{i}{\hbar}\int_0^\tau dt H_{Ii_y}(t)\} & \text{$\alpha=1$, $\beta=0$} \\
\end{cases}
\end{equation}
where $H_{Ii_x}$ and $H_{Ii_y}$ correspond to the terms in the system-environment interaction Hamiltonian $H_{Ii}$, proportional to $\sigma_x$ and $\sigma_y$ respectively.

Performing the time integrals in the exponents, we obtain
\begin{equation}
U_I(s,e,\tau)=
\begin{cases}
\exp\{i \sigma_x I(\tau)\}\times\exp\{-2 \sigma_z Q_-(\tau)\} & \text{$\alpha=0$, $\beta=1$} \\
\exp\{i \sigma_x I(\tau) - \sigma_x Q_-(\tau)\}\times\exp\{i \sigma_y Q_+(\tau)\} & \text{$\alpha=1$, $\beta=0$}
\end{cases}
\label{U_I}
\end{equation}
where
\begin{equation}
I(\tau)=\frac{1}{2}\int_0^\tau d\tau^\prime V(\tau^\prime)
\label{I_tau}
\end{equation}
\begin{equation}
Q_-(\tau)=\frac{\epsilon}{2}\sum_q (M_q a^\dagger_q - M_q^\star a_q)
\label{Q-_tau}
\end{equation}
\begin{equation}
Q_+(\tau)=\frac{\epsilon}{2}\sum_q (M_q a^\dagger_q + M_q^\star a_q)
\label{Q+_tau}
\end{equation}
and
\begin{equation}
M_q(\tau)=
\begin{cases}
\frac{g_q}{\omega_q}(1-e^{i\omega_q\tau}) & \text{$\alpha=0$, $\beta=1$} \\
\frac{g_q}{\omega_{12}-\omega_q}(1-e^{i(\omega_{12}-\omega_q)\tau}) & \text{$\alpha=1$, $\beta=0$}
\end{cases}
\end{equation}

\section{Evolution of the Reduced Density Matrix}
\label{Evolution of the Reduced Density Matrix}

To evaluate the various effects of decoherence we have to calculate the reduced density matrix of the two-level system, by tracing out the environment modes:

\begin{equation}
\rho_{ij}(s,\tau)=Tr_e\{\sum_{k,l=1}^2 \langle i|U_I P_{kl} \rho(e,0) U_I^{\dagger} |j\rangle \rho_{kl}(0)\}
\label{rho_ij}
\end{equation}
for $i,j=1,2$.

\subsection{Adiabatic case $(\alpha = 0,  \beta = 1)$}

Upon expansion of the exponentials in Eq. (\ref{U_I}), we obtain the following approximate expressions for the evolution operator and its adjoint:
\begin{equation}
U_I(s,e,\tau)=E_0+iE_x\sigma_x+E_y\sigma_y+E_z\sigma_z
\label{U_I_a}
\end{equation}
and
\begin{equation}
U_I^{\dagger}(s,e,\tau)=E_0-iE_x\sigma_x-E_y\sigma_y-E_z\sigma_z
\label{U_I-1_a}
\end{equation}
where
\begin{eqnarray}
E_0 &=& \cosh(2Q_-(\tau))\cos(I(\tau)) \\
E_x &=& \cosh(2Q_-(\tau))\sin(I(\tau)) \\
E_y &=& -\sinh(2Q_-(\tau))\sin(I(\tau)) \\
E_z &=& -\sinh(2Q_-(\tau))\cos(I(\tau)),
\end{eqnarray}
with $I(\tau)$ and $Q_-(\tau)$ defined by Eqs. (\ref{I_tau})-(\ref{Q-_tau}).


Using these expressions in Eq. (\ref{rho_ij}), we get the reduced density matrix of the two-level system:

\begin{eqnarray}
\rho_{11}&=&\rho_{11}(0)\cos^2 I + \rho_{22}(0)\sin^2 I - i[\rho_{12}(0)-\rho_{21}(0)]e^{-g_{ad}}\cos I\sin I,\label{rho_11_a} \\
\rho_{22}&=&\rho_{22}(0)\cos^2 I + \rho_{11}(0)\sin^2 I + i[\rho_{12}(0)-\rho_{21}(0)]e^{-g_{ad}}\cos I\sin I,
\label{rho_22_a} \\
\rho_{12}&=&\rho_{12}(0)e^{-g_{ad}}\cos^2 I + \rho_{21}(0)e^{-g_{ad}}\sin^2 I + i (\rho_{22}(0) - \rho_{11}(0))\cos I\sin I,
\label{rho_12_a} \\
\rho_{21}&=&\rho_{21}(0)e^{-g_{ad}}\cos^2 I + \rho_{12}(0)e^{-g_{ad}}\sin^2 I - i (\rho_{22}(0) - \rho_{11}(0))\cos I\sin I,
\label{rho_21_a}
\end{eqnarray}
\noindent
where $I=I(\tau)$, and 
\begin{equation}
g_{ad} := g_{ad}(\tau)=\gamma\int_0^\infty d\omega G(\omega)(1-\cos{\omega \tau})\coth\frac{\beta_0\omega}{2}
\end{equation}
\noindent
is the decoherence function obtained by Palma \cite{Palma96}. The dimensionless constant $\gamma$ depends on the dipole moment of the two-level system and on the Rabi frequency. The strength of the decoherence rate experienced by the two-level system is determined by $\gamma$, ie., the decoherence rate is weak for $\gamma << 1$.

The function $G(\omega)$ is the spectral function

 \begin{equation}
 G(\omega)=\omega^{n-2}e^{-\omega/\omega_c},
 \end{equation}

\noindent
where $n$ is the dimensionality of the system (usually $n=3$), and $\omega_c$ is the (usually very large) cutoff frequency, which ensures the convergence of the improper integral. In the case of phonon modes this cutoff frequency is the Debye frequency. We note that the only dependence on the two-level system is contained in $\gamma$. 

The expressions (\ref{rho_11_a})-(\ref{rho_21_a}) for the reduced density matrix elements $\rho_{ij}$ $(i,j = 1,2)$ retain the correct limit behaviors, and the trace of the density matrix is always equal to one. Indeed, in the limit of zero control we recover Palma's result \cite{Palma96}, whereby only the non-diagonal elements are affected by the decoherence process, while the diagonal elements remain unchanged.  When both the decoherence and the control are equal to zero, we recover the ideal situation in which the state remains unchanged.

\subsection{Thermal decoherence $(\alpha = 1, \beta = 0)$}

Proceeding as before, we use the approximation in Eq. (\ref{U_I}), for $\alpha=1$ and $\beta=0$, to obtain the following expressions for the evolution operator and its adjoint in the thermal decoherence case:
\begin{equation}
U_I(s,e,\tau)=E_0+iE_x\sigma_x+E_y\sigma_y+E_z\sigma_z
\end{equation}
and
\begin{equation}
U_I^{\dagger}(s,e,\tau)=E_0-iE_x\sigma_x-iE_y\sigma_y+iE_z\sigma_z
\end{equation}
where
\begin{eqnarray}
E_0 &=& \cos(I(\tau)+iQ_-(\tau))\cos(Q_+(\tau)) \\
E_x &=& \sin(I(\tau)+iQ_-(\tau))\cos(Q_+(\tau)) \\
E_y &=& \cos(I(\tau)+iQ_-(\tau))\sin(Q_+(\tau)) \\
E_z &=& \sin(I(\tau)+iQ_-(\tau))\sin(Q_+(\tau)),
\end{eqnarray}
with $I(\tau)$, $Q_-(\tau)$ and $Q_+(\tau)$ defined by Eqs.(\ref{I_tau})-(\ref{Q+_tau}). To calculate $U_I^{\dagger}(s,e,\tau)$, we have used the fact that $Q_-(\tau)$ and $Q_+(\tau)$ commute in the first order approximation for the coupling strength parameter, $\epsilon$.

The environment modes are then traced out to obtain the elements of the reduced density matrix:

\begin{eqnarray}
\rho_{11} &=& \frac{1}{2}[1 + (\rho_{11}(0)-\rho_{22}(0))e^{-2g_{th}}\cos(2I) - i(\rho_{12}(0)-\rho_{21}(0))e^{-g_{th}}\sin(2I)]
\label{rho_11_t} \\
\rho_{12} &=& [Re\{\rho_{12}(0)\}+Im\{\rho_{12}(0)\}\cos(2I)]e^{-g_{th}} - \frac{i}{2}[(\rho_{11}(0)-\rho_{22}(0))e^{-2g_{th}}\sin(2I)]
\label{rho_12_t} \\
\rho_{21} &=& [Re\{\rho_{21}(0)\}+Im\{\rho_{21}(0)\}\cos(2I)]e^{-g_{th}} + \frac{i}{2}[(\rho_{11}(0)-\rho_{22}(0))e^{-2g_{th}}\sin(2I)]
\label{rho_21_t} \\
\rho_{22} &=& \frac{1}{2}[1 - (\rho_{11}(0)-\rho_{22}(0))e^{-2g_{th}}\cos(2I) + i(\rho_{12}(0)-\rho_{21}(0))e^{-g_{th}}\sin(2I)]
\label{rho_22_t}
\end{eqnarray}

\noindent
where the decoherence function, $g_{th}$, is given by 
\begin{equation}
g_{th} := g_{th}(\tau)=\gamma\int_0^\infty d\omega \frac{1-cos[(\omega_{12}-\omega) \tau]}{(\omega_{12}-\omega)^2}\omega^3 \coth(\beta_0\omega/2)\exp(-\omega/\omega_c)
\end{equation}


In the limit of zero control, the matrix elements read:

\begin{eqnarray}
\rho_{11} &=& \frac{1}{2}(1+e^{-2g_{th}}(\rho_{11}(0) - \rho_{22}(0))) \\
\rho_{12} &=& \rho_{12}(0) e^{-g_{th}} \\
\rho_{21} &=& \rho_{21}(0) e^{-g_{th}} \\
\rho_{22} &=& \frac{1}{2}(1-e^{-2g_{th}}(\rho_{11}(0) - \rho_{22}(0)))
\end{eqnarray}
\noindent


\section{Control Strategy}
\label{Control Strategy}

Our control strategy is based on the following idea: by equating the elements of the reduced density matrix in the presence of decoherence (adiabatic or thermal) and (unknown) control (Eqs. (\ref{rho_11_a})-(\ref{rho_21_a}) or (\ref{rho_11_t})-(\ref{rho_22_t}), respectively) with the elements of the density matrix undergoing a unitary evolution, we can, in principle, determine the control that eliminates the effect of the decoherence and momentarily restores unitarity.  In our case, this simply means that the matrix elements should be restored to their initial values, i.e. the unitary evolution is the identity operator.  Indeed, since the primary goal here is to show robust elimination of decoherence, we consider that the NOT transformation has already been effected and the state is now waiting to be involved in the next operation, as specified by the ongoing quantum algorithm.  Since this ``waiting time" may be much longer that the time needed to realize the ``$\pi$-pulse" of the NOT gate, we simply ignore the latter and concentrate on the former.  Inclusion of the ``$\pi$-pulse" due to a NOT gate has been actually implemented and does not lead to any significant change other than an unnecessay complication of the formalism. This separation of the quantum gate operations from the decoherence control operations allows us to treat any one-qubit gate.

To numerically implement the control strategy, the real and imaginary parts of the four complex elements of the density matrix calculated in Section \ref{Evolution of the Reduced Density Matrix} are used in a prescribed order (see below). While this particular order is not essential, we shall define and use it consistently here, to make it easier to follow our control strategy. Indeed, since a single control cannot realize the desired effect (namely the instantaneous restoration of the ideal behavior) for all eight elements at once, we have to adjust these elements in turn. The control cycle is made up of 8 control steps, since there are 8 equations to solve for the (real and imaginary) components of the $2 \times $2 complex density matrix. This leads to a sequence of eight real transcendental equations to be solved in turn.

Ideally, the density matrix of a $2 \times 2$ system is usually completely described by three independent quantities.  It may happen that, even in the non-ideal case, for particular Hamiltonians, certain symmetries impose certain relations between the matrix elements.  However, to maintain a general and systematic character of the approach, we prefer to treat them all as independent variables, at the expense of a slight lengthening of the control process.  Since this process eventually stabilizes into a rather short cycle, we do not consider this to be a serious drawback.

The algorithm is applied identically for either the adiabatic or the thermal case; thus, we describe it for a generic decoherence function, denoted $g$. In the first cycle of eight time steps, we start by considering $\rho_{11R}(1):=\rho_{11R}(g(1),I(1))$ after one time step. To determine the control pulse required to set $\rho_{11R}(1)$ equal to its unitary value after applying a NOT gate, we find a control value $I(1)$ which solves the equation $\rho_{11R}(1) = \rho_{11R}(0)$.

Finding the appropriate solutions of the above transcendental equation requires a certain care.  Indeed, if there are no solutions in a given interval, it is impossible to restore the unitary behavior.  On the other hand, if there are multiple solutions, it becomes very difficult to guarantee that the same control is used for every cycle.  To obtain a unique and consistent solution, we had to choose carefully a specific interval.  Of course, this caveat applies for the solutions of the other equations as well (see below).

Denoting the solution of this equation by $I(1)$, we reset $\rho_{11R}(1)$ exactly to its original value by using a control pulse $I(1)$. After the first time step, all the other elements will have suffered the effect of decoherence, $g(1)$, and the effect of the control pulse, $I(1)$:
\begin{eqnarray}
\rho_{11R}(1) &=& \rho_{11R}(g(1),I(1)) = \rho_{11R}(0) \\
\rho_{11I}(1) &=& \rho_{11I}(g(1),I(1)) \\
\rho_{12R}(1) &=& \rho_{12R}(g(1),I(1)) \\
\rho_{12I}(1) &=& \rho_{12I}(g(1),I(1)) \\
\rho_{21R}(1) &=& \rho_{21R}(g(1),I(1)) \\
\rho_{21I}(1) &=& \rho_{21I}(g(1),I(1)) \\
\rho_{22R}(1) &=& \rho_{22R}(g(1),I(1)) \\
\rho_{22I}(1) &=& \rho_{22I}(g(1),I(1))
\end{eqnarray}

By definition, the restoration of unitarity executed by the control $I$ is exact.

The control strategy advances through the next seven time steps in a similar fashion. The required control pulses are determined by solving each of the other seven density matrix equations in the order presented above.

For example, at the second time step, we consider the equation for $\rho_{11I}(2)=\rho_{11I}(g(2),I(1)+I(2))$. To determine the control pulse required to set $\rho_{11I}(2)$ equal to its desired value, we find the control $I(2)$ which solves the equation $\rho_{11I}(2) = \rho_{11I}(0)$. After applying this control, the values of the matrix elements after the second time step are given by:
\begin{eqnarray}
\rho_{11R}(2) &=& \rho_{11R}(g(1),I(2)) \\
\rho_{11I}(2) &=& \rho_{11I}(g(2),I(1)+I(2)) = \rho_{11I}(0) \\
\rho_{12R}(2) &=& \rho_{12R}(g(2),I(1)+I(2)) \\
\rho_{12I}(2) &=& \rho_{12I}(g(2),I(1)+I(2)) \\
\rho_{21R}(2) &=& \rho_{21R}(g(2),I(1)+I(2)) \\
\rho_{21I}(2) &=& \rho_{21I}(g(2),I(1)+I(2)) \\
\rho_{22R}(2) &=& \rho_{22R}(g(2),I(1)+I(2)) \\
\rho_{22I}(2) &=& \rho_{22I}(g(2),I(1)+I(2))
\end{eqnarray}

Note that the decoherence function used to calculate $\rho_{11R}(2)$ is different from those for all the other matrix elements, since $\rho_{11R}(1)$ was reset to $\rho_{11R}(0)$ after the first time step. In general, resetting any one of the matrix elements to its corresponding original value implies that the cumulative effect of the control pulses applied since the last correction (eight time steps in the past) cancels out the effect of decoherence since the last correction. 

To complete the first cycle we proceed to solve the equations for the next six matrix elements,  to determine the required control pulses. By the end of the first cycle of eight time steps each of the matrix elements depends on different decoherence and control values:
\begin{eqnarray}
\rho_{11R}(8) &=& \rho_{11R}(g(7),I(2)+I(3)+I(4)+I(5)+I(6)+I(7)+I(8)) \\
\rho_{11I}(8) &=& \rho_{11I}(g(6),I(3)+I(4)+I(5)+I(6)+I(7)+I(8)) \\
\rho_{12R}(8) &=& \rho_{12R}(g(5),I(4)+I(5)+I(6)+I(7)+I(8)) \\
\rho_{12I}(8) &=& \rho_{12I}(g(4),I(5)+I(6)+I(7)+I(8)) \\
\rho_{21R}(8) &=& \rho_{21R}(g(3),I(6)+I(7)+I(8)) \\
\rho_{21I}(8) &=& \rho_{21I}(g(2),I(7)+I(8)) \\
\rho_{22R}(8) &=& \rho_{22R}(g(1),I(8)) \\
\rho_{22I}(8) &=& \rho_{22I}(g(8),I(1)+I(2)+I(3)+I(4)+I(5)+I(6)+I(7)+I(8)) = \rho_{22I}(0) 
\end{eqnarray}

\noindent
a pattern which will recur in every cycle of eight time steps.

After one time step in the second cycle of eight time steps, we are back at the starting point.  To determine the control pulse required to set $\rho_{11R}(9)$ equal to its initial value, we find a control value $I(9)$ which solves the equation $\rho_{11R}(g(8), I(2)+I(3)+I(4)+I(5)+I(6)+I(7)+I(8)+I(9)))=\rho_{11R}(g(1), I(1))=\rho_{11R}(0)$.

In the following time step, set $\rho_{11I}(10)=\rho_{11I}(g(8),I(3)+I(4)+I(5)+I(6)+I(7)+I(8)+I(9)+I(10))$, and so on.


This procedure is repeated as long as the quantum state of the two-level system has to be maintained, i.e. the ``waiting time" mentioned previously until the next logic gate is applied.  We note two important things for the applicability of the scheme.  First, the knowledge of the decoherence function is needed only for a finite period of time (in the example above only eight time steps).  Second, after initial transients, controls will stabilize and the whole control cycle will repeat itself periodically.  In other words, for each sequence the controls can be calculated off-line and applied in the required order. This behavior has indeed been observed: the values of the control pulses were rapidly stabilized after the first cycle of 8 time steps, as shown in Figures 1 and 2. The stabilization time depends on initial conditions, and the magnitude of the decoherence.


\section{Discussion of the Results}
\label{Discussion of the Results}

The graphs shown in Figs. 1 and 2 illustrate the typical evolution of the matrix elements while applying this control strategy in both adiabatic and thermal decoherence situations.


As  expected, the relative size of the time step between control pulses determines the amount of deviation of the matrix elements from their unitary values.  Also, the size and frequency of the control needed to restore perfectly the ideal situation depends on the strength of the decoherence.  For $\gamma=1$ (extremely strong decoherence), the frequency of the control can be decreased only to about twice the Rabi frequency.  If we decrease the frequency even more, we cannot restore exact unitarity, at least not by using this scheme. Since the control algorithm described here is based on periodic perfect restoration of the ideal situation, we shall not discuss this case any further.  However, imperfect restoration of unitarity is likely to happen and is $\it not$ at all hopeless: this situation will be analyzed in future work.  As the strength of the decoherence, $\gamma$, decreases (strong to medium decoherence), we can also decrease the frequency of the control pulses while keeping the same amount of deviation of the matrix elements from their unitary values. Of course, higher frequencies result in better restorations of unitarity.

The same observations apply for thermal decoherence.  Since the latter affects all elements of the density matrix, it is reasonable to expect that the same amount of control will be less efficient here than in the adiabatic situation.  This is indeed the case, as illustrated in Fig. 2.

If the controls are perturbed by noise, the control is still very efective. It is important to note that the amplitude of the noise is calibrated with respect to the state and not with respect to the amplitude of the control needed in the noiseless situation, which in some cases is extremely small. 

\begin{figure}[tp]
\label{Fig1}
(a)\scalebox{0.25}{\includegraphics{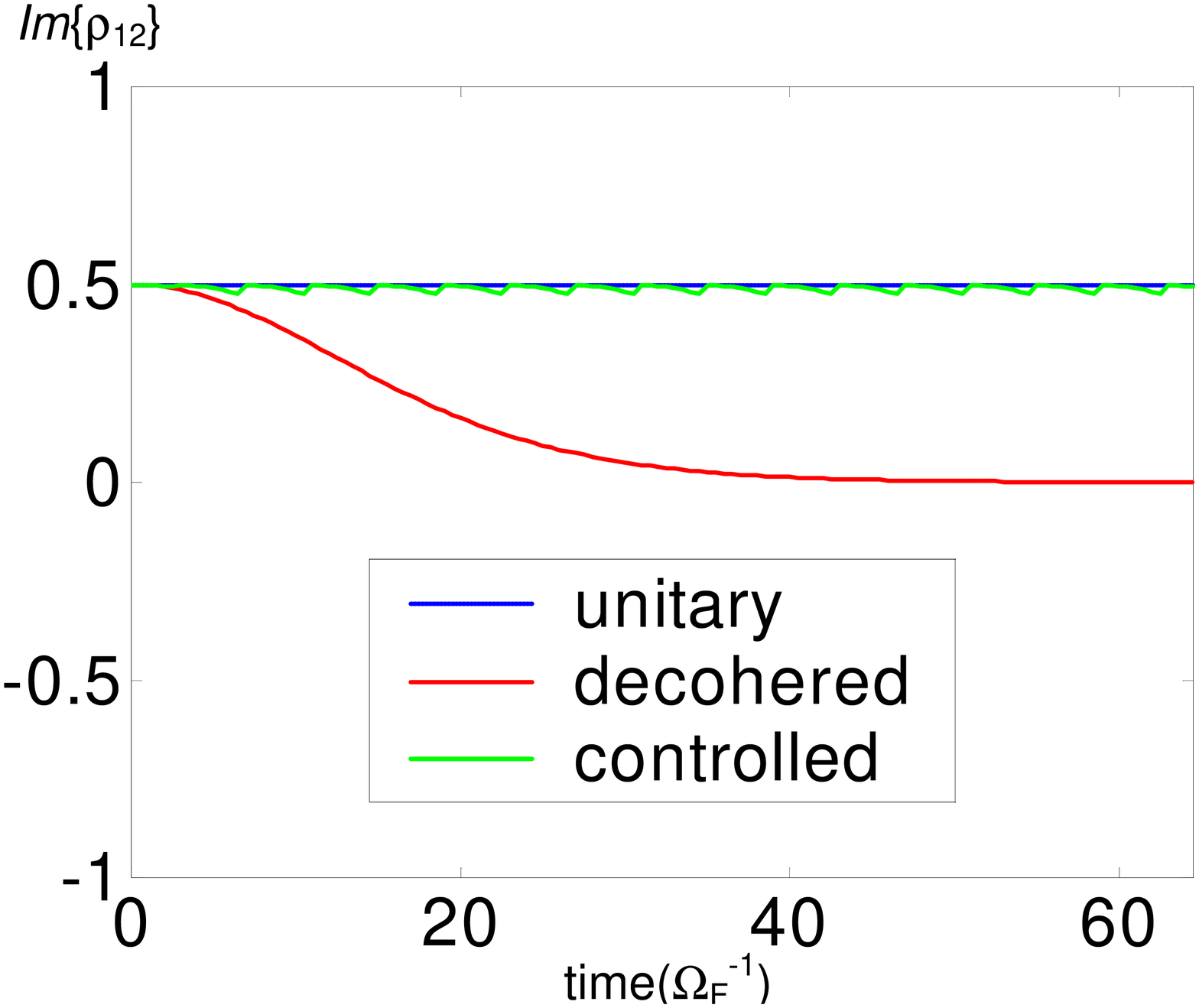}}(b)\scalebox{0.25}{\includegraphics{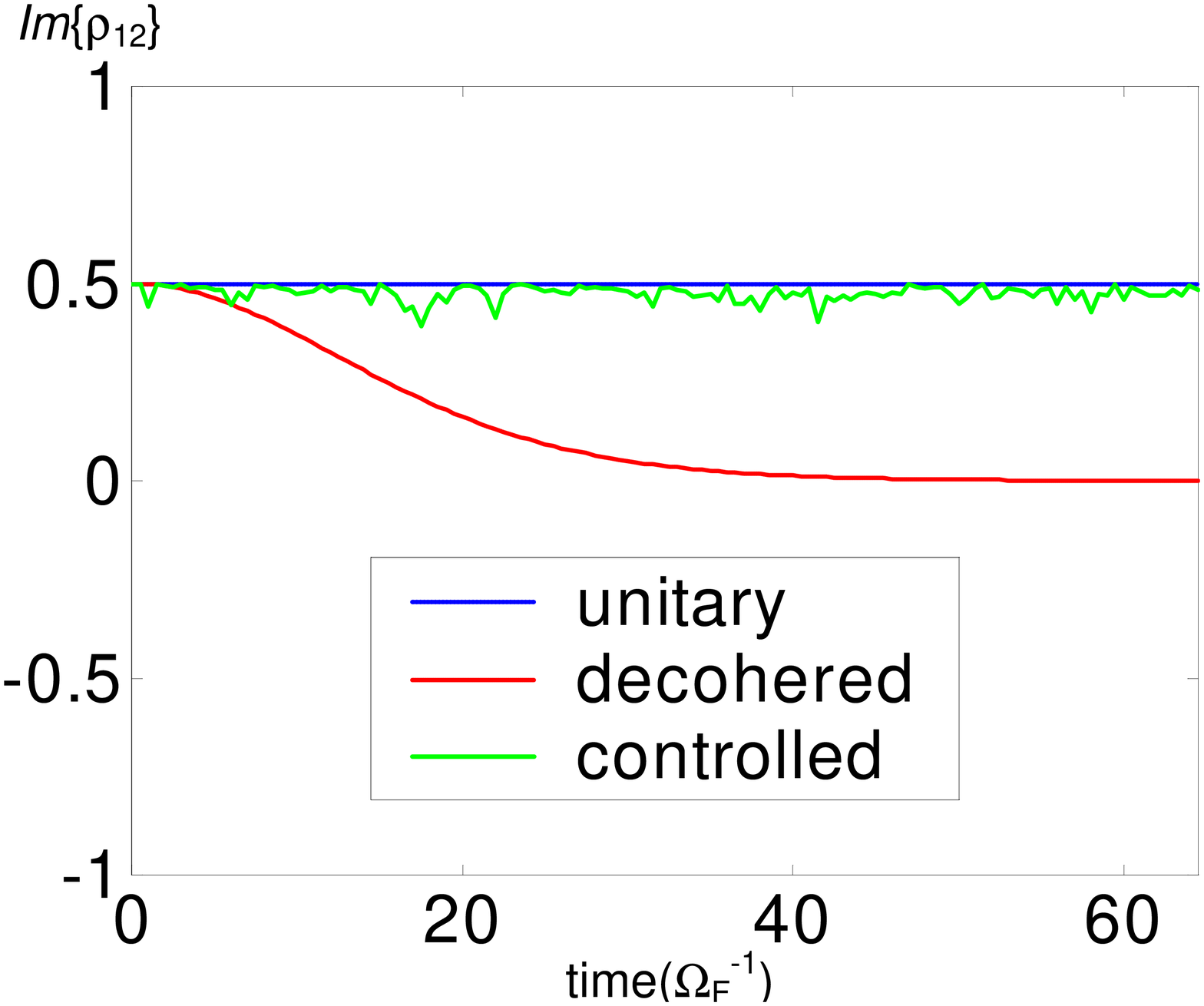}}
(c)\scalebox{0.25}{\includegraphics{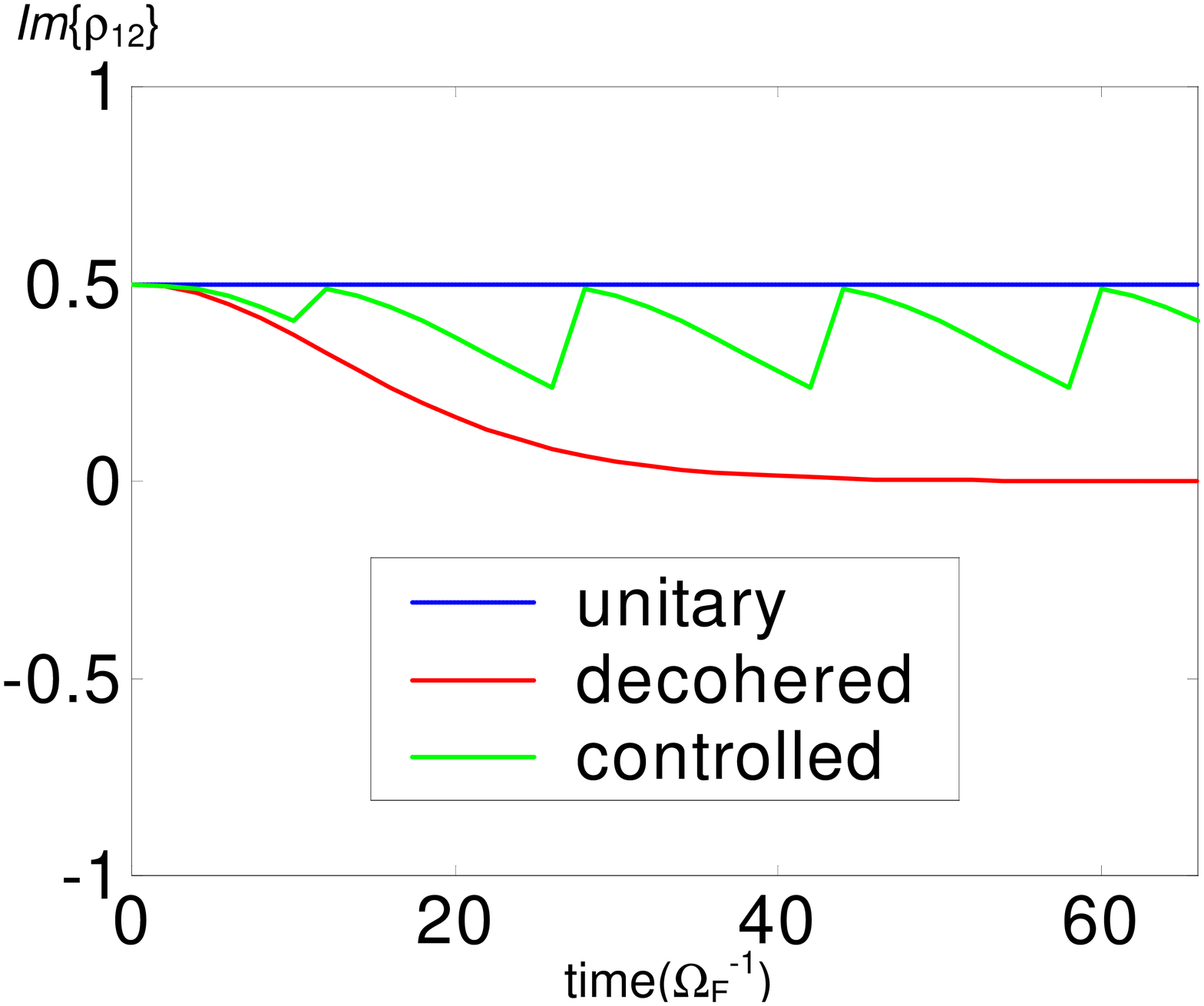}}(d)\scalebox{0.25}{\includegraphics{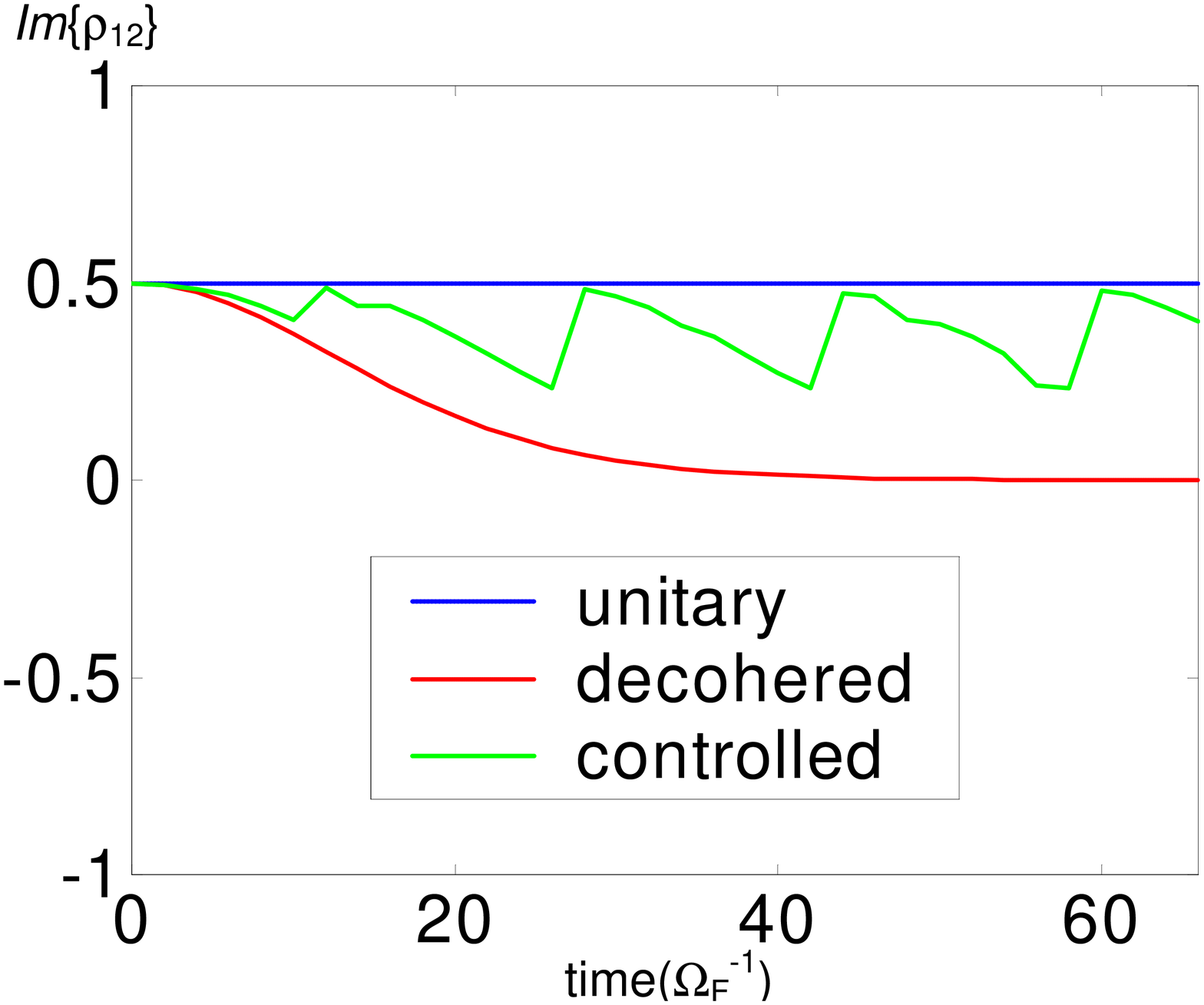}}
\caption{The unitary, adiabatically decohered, and controlled evolution of the $Im\{\rho_{12}\}$ element of the density matrix for the initial state $\frac{i}{\sqrt2}|1> + \frac{1}{\sqrt2}|2>$. The dimensionless decoherence rate is set to strong, $\gamma=1$, and the other two parameters characterizing these plots are: the time between control pulses, $T$ (scaled in terms of the Rabi frequency), and the standard deviation, $\Delta I$, of the control pulses after adding normally-distributed noise. (a) $T=0.5$, $\Delta I=0$; (b) $T=0.5$, $\Delta I=0.1$; (c) $T=2$, $\Delta I=0$; (d) $T=2$, $\Delta I=0.1$. }
\end{figure}

\begin{figure}[tp]
\label{Fig2}
(a)\scalebox{0.25}{\includegraphics{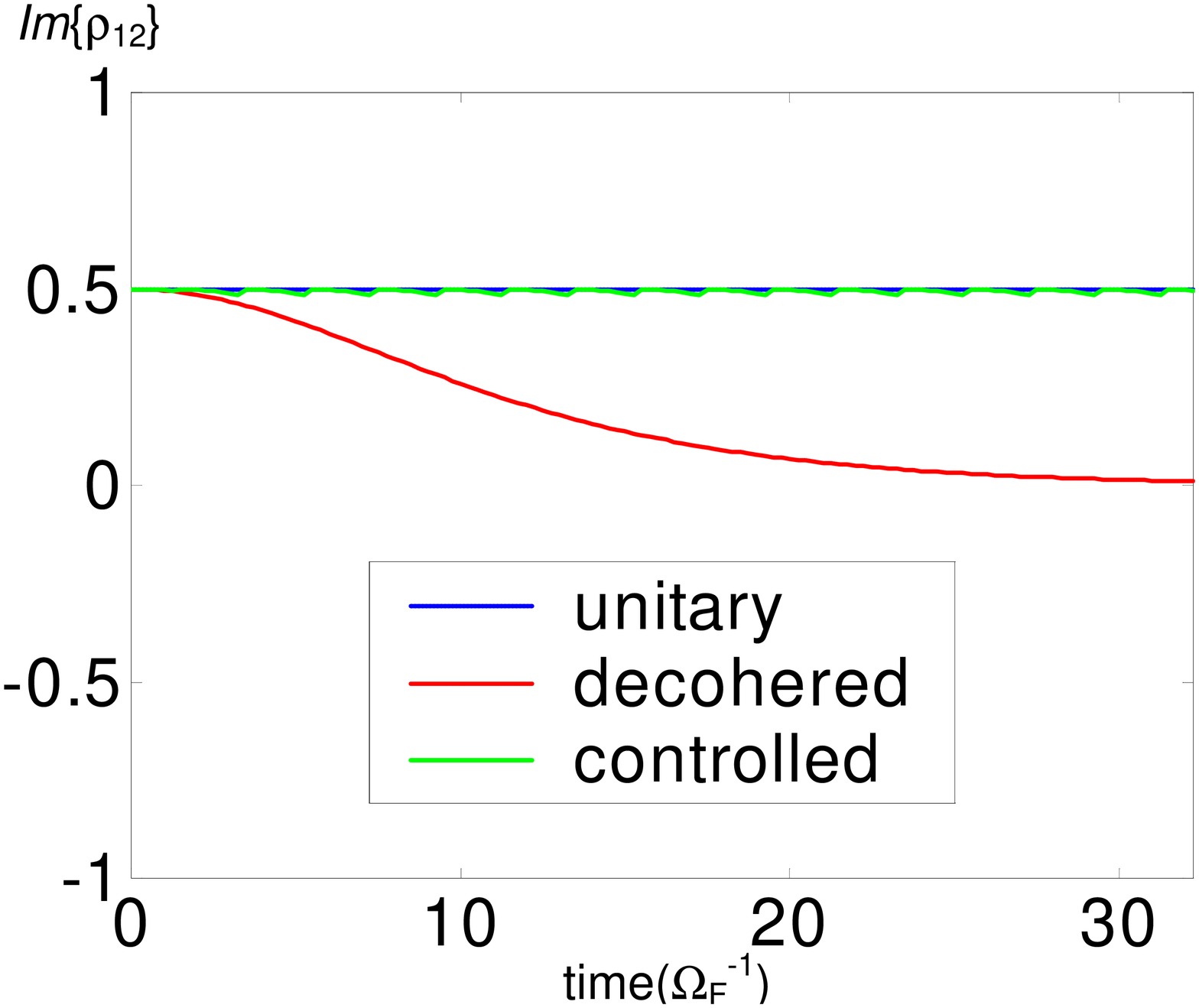}}(b)\scalebox{0.25}{\includegraphics{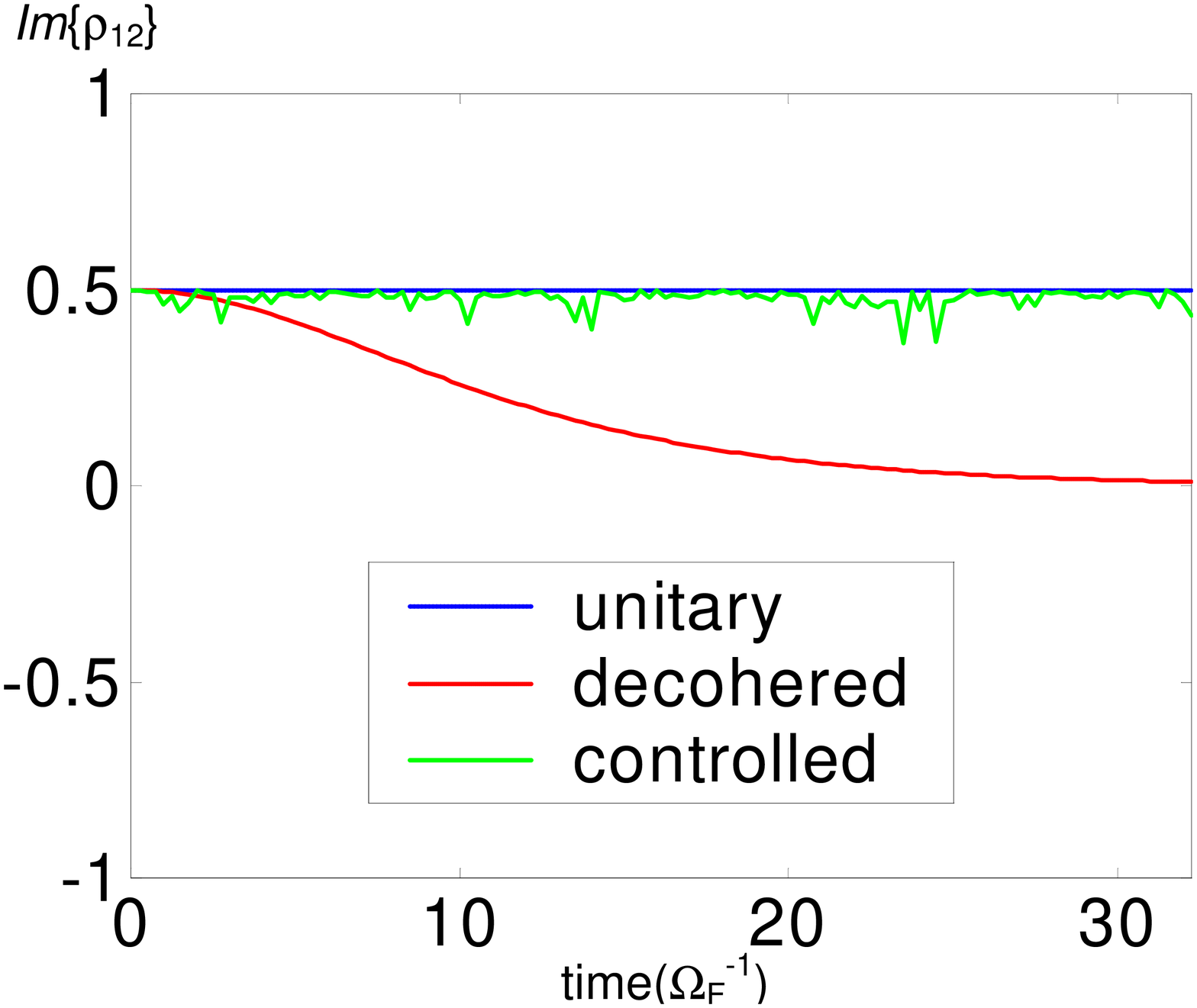}}
(c)\scalebox{0.25}{\includegraphics{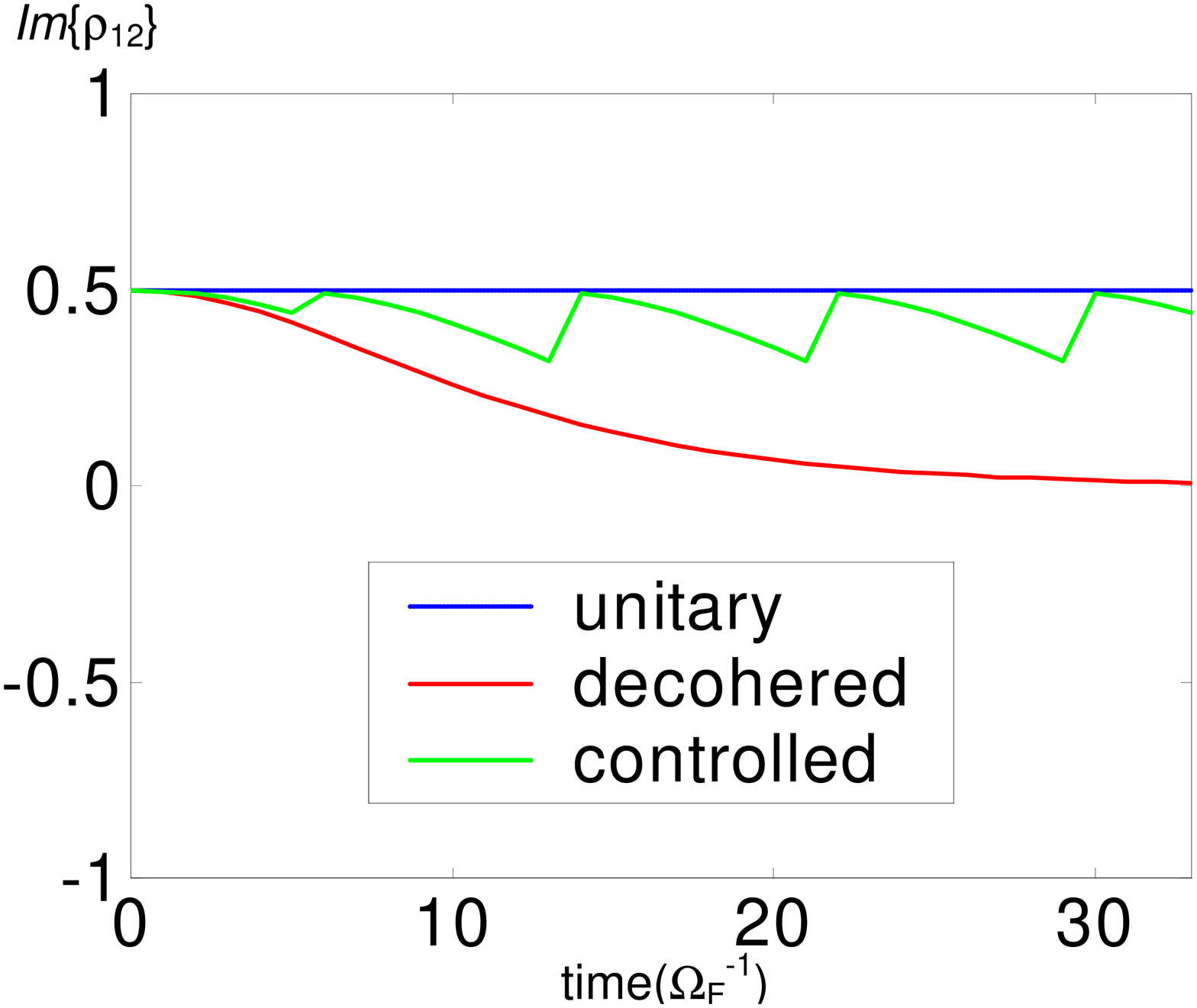}}(d)\scalebox{0.25}{\includegraphics{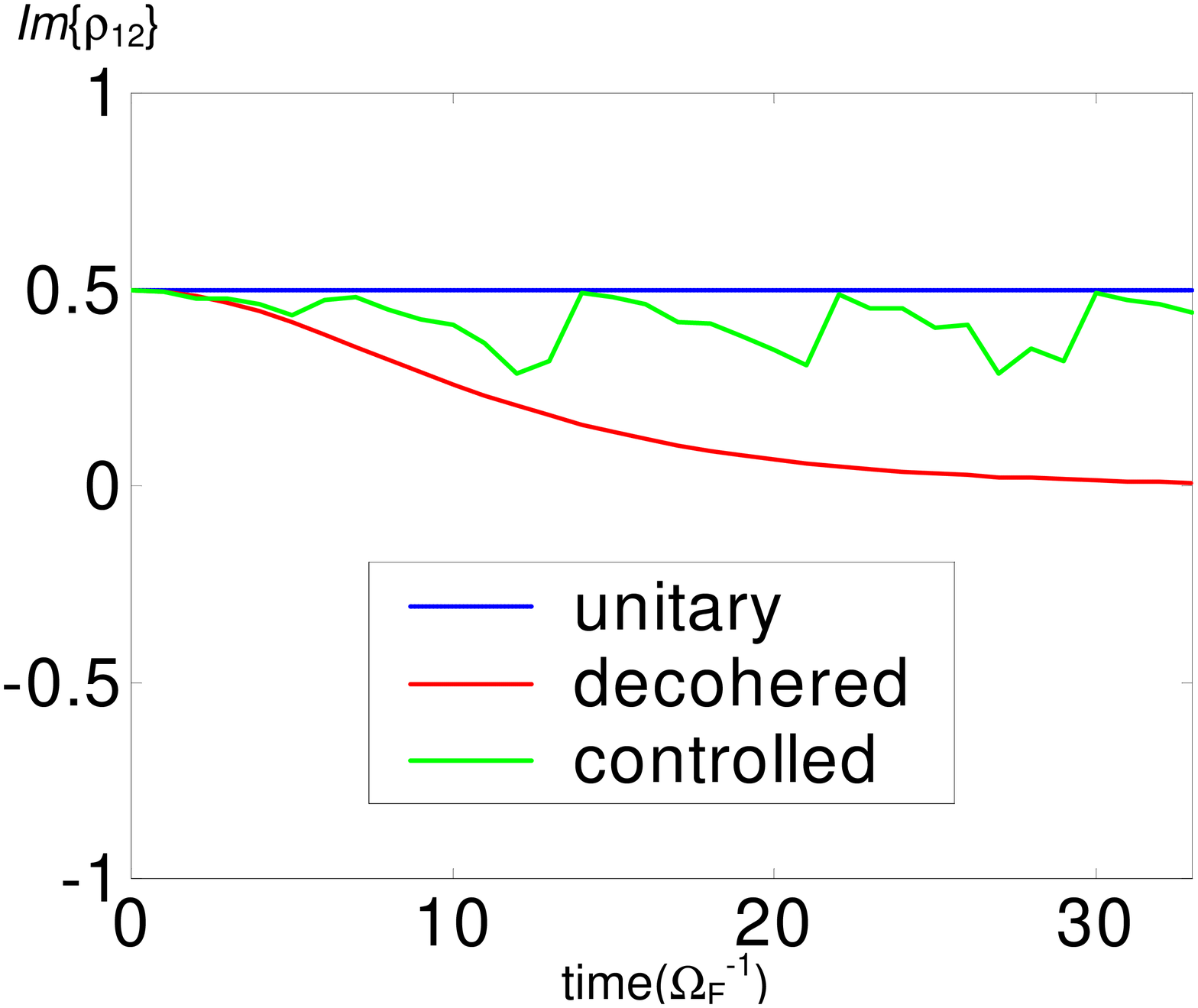}}
\caption{The unitary, thermally decohered, and controlled evolution of the $Im\{\rho_{12}\}$ element of the density matrix for the initial state $\frac{i}{\sqrt2}|1> + \frac{1}{\sqrt2}|2>$.  The dimensionless decoherence rate is set to strong, $\gamma=1$, and the other two parameters characterizing these plots are: the time between control pulses, $T$ (scaled in terms of the Rabi frequency), the decoherence rate, $\gamma$ (dimensionless), and the standard deviation, $\Delta I$, of the control pulses after adding normally-distributed noise. (a) T=0.25, $\Delta I=0$; (b) $T=0.25$, $\Delta I=0.1$; (c) $T=1$, $\Delta I=0$; (d) $T=1$, $\Delta I=0.1$. }
\end{figure}

In conclusion, we presented an open loop control scheme for quantum gates which is: (i) realistic - the control and decoherence act simultaneously, and the model is able to deal with adiabatic as well as thermal decoherence; 
(ii) relatively simple to calculate, (iii) efficient to implement, and (iv) robust to noise both in the decoherence functions as well as in the control pulses.

The results presented in this paper depend, among other factors, on the model Hamiltonian, and the approximation used in the tracing out 
procedure. However, modifications of either would not alter the main features of the proposed control strategy.

While our control scheme shares several common features with the quantum feedback scheme proposed by Wang \textit{et al.} \cite{Wang01} and the quantum bang-bang control, there are also significant differences between our scheme and these approaches, which we briefly discuss here.

The quantum feedback eliminates specifically the decoherence effects of spontaneous emission, whereas the present scheme is more general. Both the feedback scheme and the approach proposed here aim to restore (at least approximately) the unitarity i.e., maintain the information contained in the quantum state. However our scheme assumes prior knowledge of the initial state in order to maintain it unchanged, whereas the quantum feedback scheme drives the state to a target state regardless of initial conditions. Finally, our scheme does not require any measurements on the two-level system or the environment, and does not involve any feedback.

Similarly to quantum bang-bang control schemes \cite{Tombesi_ol,Viola98,Viola99a,Viola99b}, the control proposed here relies on a series of relatively fast control pulses to dynamically cancel the system-environment interaction.  However, in our approach, we tailor the control according to the specific nature of decoherence present in a physical system.  This has certain advantages, namely: (i) controls are applied only at the required frequency and magnitude, without overtaxing the available resources; (ii) the limits of applicability are also clear, beyond which a specific open loop scheme would not be able to restore exact unitarity; (iii) since the environment is not supposed to change significantly during the quantum computation, the control sequence stabilizes very quickly to a set of periodic controls, as shown in Figs. 1 and 2 (this implies that in most situations of interest the controls can be computed off-line and then simply and efficiently applied in the prescribed sequence); and (iv) robustness to noise appears quite strong, both in the adiabatic and thermal cases.

The main difference between our scheme and the bang-bang control scheme proposed by Viola and Lloyd \cite{Viola98} is that, on the one hand we can handle the more difficult case involving thermal decoherence, and on the other hand the control pulses are tailored to the task instead of being generic. Indeed, a cycle of two predetermined $\pi/2$-pulses is used in Viola and Lloyd's scheme for adiabatic decoherence. The evolution of these pulses is disjoint from the evolution of the system and the environment, in order to ensure that the cycle evolution can be simplified to an expression containing only $\sigma_z$. However, the control proposed in Ref. \cite{Viola98} cannot be used as such in the case of thermal decoherence, which has a system-bath interaction Hamiltonian proportional to $\sigma_x$, even if the control Hamiltonian is taken to be a linear combination of $\sigma_x$ and $\sigma_y$. On the contrary, in our scheme the evolution of the pulses is \textit{not} disjoint from the evolution of the system and the environment, which allows us to handle a general combination of thermal and adiabatic decoherence by employing the BCH approximation. Moreover our controls do not have \textit{a priori} assigned values. In general, the stabilized pulses have a much smaller amplitude than the $\pi/2$-pulses, for the same pulse duration, and they are applied at a slower rate. In principle, this improves the efficiency of the control.

The limitation of our scheme is its reliance on knowledge of the target state that we want to maintain, e.g., the initial values of the reduced density matrix of the two-level system. Since the target state changes during the course of the quantum computation, the applicability of this control scheme is presently limited to short calculations. We intend to explore this aspect together with different Hamiltonians, approximations, and control schemes in the future.

\section* {Acknowledgments}
\label {Acknowledgments}
{This research was supported in part by the U.S. Department of Energy, Office of Basic Energy Sciences.  The Oak Ridge National Laboratory is managed for the U.S. DOE by UT-Battelle, LLC, under contract No. DE-AC05-00OR22725. The authors thank P. Tombesi for pointing out several important references. We are particularly indebted to L. Viola for several pertinent and insightful comments, suggestions, and questions that led to the clarification and improvement of this paper.}


\begin{thebibliography}{99}
\bibitem{Pitt}Pittenger A O 2000 {\it An Introduction to Quantum Computing Algorithms} (Boston: Birkh\"auser)
\bibitem{NC}Nielsen M A and Chuang I L 2000 {\it Quantum Computation and Quantum Information} (Cambridge: Cambridge University Press)
\bibitem{Gardiner91}Gardiner C W 1991 {\it Quantum Noise} (Berlin: Springer)
\bibitem{Giulini}Giulini D, Joos E, Kiefer C, Kupsch J, Stamatescu I -O and Zeh H D 1996 {\it Decoherence and the Appearance of a Classical World in Quantum Theory} (Berlin: Springer)
\bibitem{BEZ}Bouwmeester D, Ekert A and Zeilinger A (Eds.) 2000 {\it The Physics of Quantum Information} (Berlin: Springer)
\bibitem{Weiss99}Weiss U 1999 {\it Quantum Dissipative systems} (Singapore: World Scientific)
\bibitem{Tombesi_ol}Vitali D and Tombesi P 1999 Phys. Rev. A {\bf 59} 4178; Vitali D and Tombesi P 2002 Phys. Rev. A {\bf 65} 012305; Mancini S, Vitali D, Bonifacio R, and Tombesi P 2001 {\it Preprint} quant-ph/0108011
\bibitem{Viola98}Viola L and Lloyd S. 1998 Phys. Rev. A {\bf 58}(4) 2733
\bibitem{Viola99a}Viola L, Knill E and Lloyd S 1999 Phys. Rev. Lett. {\bf 82}(12) 2417
\bibitem{Viola99b}Viola L, Lloyd S and Knill E 1999 Phys. Rev. Lett. {\bf 83}(23) 4888
\bibitem{Zanardi97}Zanardi P and Rasetti M 1997 Phys. Rev. Lett. {\bf 79} 3306 
\bibitem{Lidar98}Lidar D A, Chuang I L and Whaley K B 1999 Phys. Rev. Lett. {\bf 81} 2594
\bibitem{Preskill98}Preskill J 1998 Proc. Roy. Soc. Lond. A {\bf 454} 385
\bibitem{Knill00}Knill E, Laflamme R and Viola L 2000 Phys. Rev. Lett. {\bf 84} 2525
\bibitem{Tombesi_qf}Tombesi P and Vitali D 1995 Phys. Rev. A {\bf 51} 4913; Goetsch P, Tombesi P and Vitali D 1996 Phys. Rev. A {\bf 54} 4519; Tombesi P and Vitali D 1995 Appl. Phys. B {\bf 60} S69; Tombesi P, Vitali D and Milburn G J 1997 Phys. Rev. Lett. {\bf 79} 2442; Tombesi P, Vitali D and Milburn G J 1998 Phys. Rev. A {\bf 57} 4930; Fortunato M, Raimond J M, Tombesi P and Vitali D 1999 Phys. Rev. A {\bf 60}, 1687
\bibitem{Wang01}Wang J, Wiseman H and Milburn G J 2001 J. Chem. Phys. {\bf 268} 221
\bibitem{Viola00}Viola L, Knill E and Lloyd S 2000 Phys. Rev. Lett. {\bf 85} 3520
\bibitem{WL}Wu L A and Lidar D A 2001 {\it Preprint} quant-ph/0112144
\bibitem{Kwiat00}Kwiat P G, Berglund A J, Altepeter J B and White A G 2000 Science {\bf 290} 498
\bibitem{Kielpinski01}Kielpinski D, Ben-Kish A, Britton J, Meyer V, Rowe M A, Sackett C A, Itano W M, Monroe C and Wineland D J 2001 Science {\bf 291}, 1013
\bibitem{Viola01}Viola L, Fortunato E M, Pravia M A, Knill E, Laflamme R and Cory D G 2001 Science {\bf 293}, 2059
\bibitem{Knill96}Knill E, Laflamme R and Zurek W 1998 Science {\bf 279}, 342
\bibitem{Leggett87}Leggett A J, Chakravarty S, Dorsey A T, Fisher M P A, Garg A and Zwerger W 1987 Rev. Mod. Phys. {\bf 59}, 1
\bibitem{Kampen95}van Kampen N G 1995 J. Stat. Phys. {\bf 78} 299
\bibitem{Palma96}Palma G M, Suominen K -A and Ekert A K 1996 Proc. R. Soc. London A {\bf 452} 567
\bibitem{Mozyrsky98}Mozyrsky D and Privman V 1998 J. Stat. Phys. {\bf 91} 787
\bibitem{ZHP}Zurek W H, Habib S and Paz J P 1993 Phys. Rev. Lett. {\bf 70}(9) 1187
\bibitem{Scully_qo} Scully M S and Zubairy M S 1996 {\it Quantum Optics} (Cambridge: Cambridge University Press)
\bibitem{Gardiner_qn} Gardiner C W and Zoller P 2000 {\it Quantum Noise} (Berlin: Springer)
\end{thebibliography}
\end{document}